# Towards On-Chip Optical FFTs for Convolutional Neural Networks

Jonathan K. George  
Department of Electrical and Computer Engineering  
The George Washington University  
Washington, D.C., USA  
jkg0@gwmail.gwu.edu

Hani Nejadriahi  
Department of Electrical and Computer Engineering  
The George Washington University  
Washington, D.C., USA  
hani_nejariahi@gwmail.gwu.edu

Volker J. Sorger  
Department of Electrical and Computer Engineering  
The George Washington University  
Washington, D.C., USA  
sorger@gwu.edu

*Abstract*— Convolutional neural networks have become an essential element of spatial deep learning systems. In the prevailing architecture, the convolution operation is performed with Fast Fourier Transforms (FFT) electronically in GPUs. The parallelism of GPUs provides an efficiency over CPUs, however both approaches being electronic are bound by the speed and power limits of the interconnect delay inside the circuits. Here we present a silicon photonics based architecture for convolutional neural networks that harnesses the phase property of light to perform FFTs efficiently. Our all-optical FFT is based on nested Mach-Zender Interferometers, directional couplers, and phase shifters, with backend electro-optic modulators for sampling. The FFT delay depends only on the propagation delay of the optical signal through the silicon photonics structures. Designing and analyzing the performance of a convolutional neural network deployed with our on-chip optical FFT, we find dramatic improvements by up to $10^4$ when compared to state-of-the-art GPUs when exploring a compounded figure-of-merit given by power per convolution over area. At a high level, this performance is enabled by mapping the desired mathematical function, an FFT, synergistically onto hardware, in this case optical delay interferometers.

## I. Introduction

One singular attribute of the electromagnetic wave is its ability to perform addition and subtraction as it propagates. These arithmetic operations, commonly known as wave interference, are the basis of holography, phased array antennas, and interferometric microscopy. The only energy consumed by these arithmetic operations is the loss incurred by the propagation of the wave. One of the earliest uses for optics in computing was frequency domain filtering with Fourier optics. In these systems, a lens is used to convert an image into the frequency domain where filtering can occur with the result being converted back into the spatial domain with a second lens. Operating on the full image, these systems are highly parallel but also bulky. The idea of using optical interference for the Fast Fourier Transform (FFT) was first introduced by Marhic [1]. In this Optical FFT (OFFT) system, star couplers are used to perform addition and subtraction and length differences are used to rotate phase. Advances in on-chip photonics as well as simplifications of the waveguide layout [2]-[4] have led to realizable OFFTs in Silicon-On-Insulator (SOI) technology. While the most immediate application for the OFFT is in high bandwidth communications where frequency domain representation can break a large bandwidth into many sub-bands for Orthogonal Frequency Division Multiplexing (OFDM), a more recent application for frequency domain representation is found in Convolutional Neural Networks (CNNs).

A CNN is neural network where instead of fully connecting each input to each output with weights, convolutional filtering connects the network in a spatially local manner [5]. This convolutional filtering is normally performed by Graphics Processing Units (GPUs). The GPUs convert the input data to the frequency domain with a forward FFT where it is multiplied by a kernel and then converted back into the spatial domain with an inverse FFT. By using an OFFTs for convolution instead GPUs, a system can be built to take advantage of the energy efficient arithmetic of wave interference to perform the convolutions of the CNN (Fig. 1).

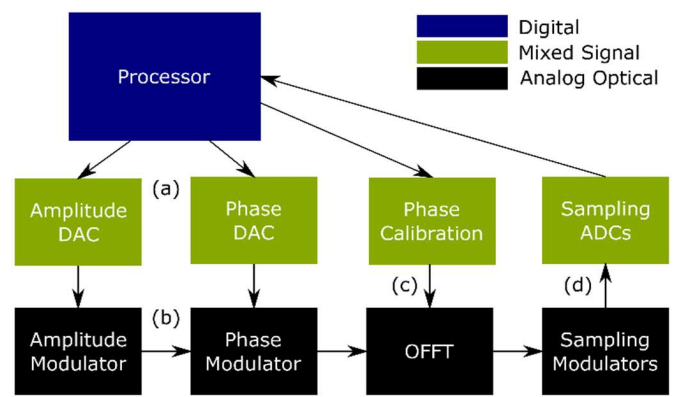

Fig. 1. Block diagram of the on-chip OFFT showing the data flow from the processor to the amplitude and phase Digital to Analog Converters (DACs) (a), being modulated onto an optical carrier (b) flowing through the phase calibrated OFFT network (c) and being converted back into the digital domain (d) with sampling Analog to Digital Converters (ADCs) and optional sampling modulators.

## II. THE OPTICAL COOLEY-TUKEY FFT

The OFFT is built on three passive components: the 2 x 2 coupler is used for addition and subtraction, waveguides with short path differences are used for phase rotation, and waveguides with long path differences are used for signal delay [1], [2]. While in principal an OFFT network could be created with perfect phase alignment at a specific temperature, in practice active phase calibration is required to compensate for fabrication and temperature variance. This phase calibration is normally accomplished with heating elements placed along one of the waveguide paths of each waveguide pair.

The Cooley-Tukey FFT requires two operations: addition and multiplication by a phase. The 2 x 2 optical coupler forms the principal addition equation of the OFFT, Eq. 1.

$$\beta_1 = \frac{1}{\sqrt{2}}(-\alpha_1 + \alpha_2)$$
$$\beta_2 = \frac{1}{\sqrt{2}}(\alpha_1 + \alpha_2) \quad (1)$$

Where $\beta_1$ and $\beta_2$ are the outputs and $\alpha_1$ and $\alpha_2$ are the outputs of the 2 x 2 coupler. The phase multiplication required by the Cooley-Tukey FFT can be implemented optically by phase difference, Eq. 2.

$$\epsilon_{xy} = exp(-i2\pi xy/N) \quad (2)$$

With these two components, the butterfly pattern (Fig. 2) of the Cooley-Tukey FFT can be built using only passive optics.

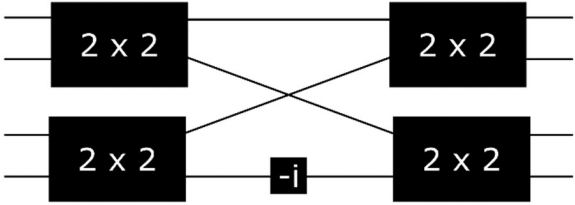

Fig. 2. The 4 x 4 butterfly pattern of the Cooley-Tukey OFFT is composed of a set of passive optical on-chip components in SOI technology. 2 x 2 optical couplers provide addition and subtraction while a small path-length difference in one of the waveguide branches provides the reciprocal root-of-unity change in phase.

## III. OPTICAL FFT AND SOI FABRICATION

The OFFT in silicon photonics becomes a network of delay waveguides, Y-branches, Mach Zehnder Interferometers (MZIs), and heater-calibrated phase delay waveguide segments (Fig. 3). Delay is implemented with spiral delay stages. The spirals scale in area proportional to their length. The length of the first spiral is the greatest and they diminish in length with $(1/2)^k$, where k is the stage index. There are $log_2(N)$ delay stages each with $2^k$ spirals. Even though the number of spirals doubles with each stage, the area stays constant due to the spiral length halving with each stage. Thus, the area relative to the first spiral scales with $log_2(N)$ and the first spiral with scales with N for a total area scaling of $Nlog_2(N)$.

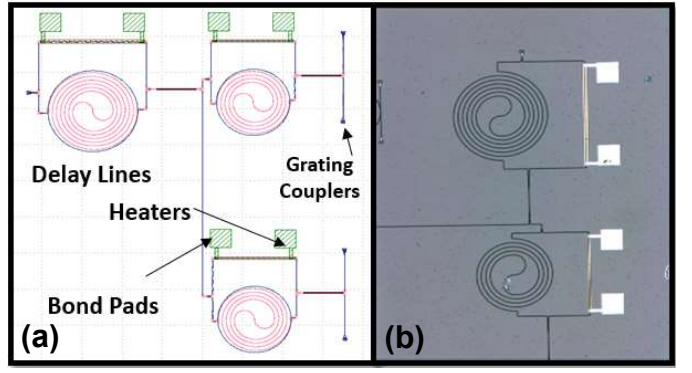

Fig. 3. An N = 4 OFFT in silicon photonics, layout (a) and microscope image (b), shows the area budget being dominated by spiral delay waveguide stages. Physical area of the optical delay required in the serial architecture limits the scaling performance of the OFFT to small N.

The second most significant scaling factor in the OFFT is found in the increasing optical losses from the growing number of branches and waveguide delay lengths at high N. The optical power at each output must be at least enough to meet the noise requirements for the number of bits at the detector. At high N the optical loss becomes dominated by the $Nlog_2 N$ scaling of spiral length.

## IV. 2D OPTICAL FFT

Unlike OFDM, convolution works with spatial data and hence requires the 2D FFT. The M x N 2D FFT can be composed from row and column operations of M length N FFTs plus N length M FFTs. For a square matrix this becomes 2N FFTs of length N. In the OFFT there is a choice between implementing a large 2D FFT network directly or implementing a smaller 1D FFT in the complex domain and using it repeatedly in time for each row and column operation [6]. While complex OFFTs have not been directly discussed in literature, Eq. 1 and Eq. 2 hold for complex signals. To generate a complex OFFT an additional reference signal path must be mixed with the output prior to digitization to determine both phase and amplitude, just as in an optical heterodyne quadrature phased shift keying (QPSK) receiver. Alternatively, the phase can be measured by phase-difference as in an optical differential phased shift keying (DPSK) receiver. In the phase-difference method, two cycles of the OFFT are required. The first cycle sends a known signal through the OFFT network and measure the real and imaginary part at the output. The second cycle sends the actual signal through the network and measures the real and imaginary part of the signal relative to the calibration signal. In either the heterodyne or the differential case, the complex and real measurements can be split into two separate cycles of the OFFT to conserve ADCs. In the two-cycle approach, the unmodified signal is fed through the OFFT and the real part of the output is measured. Next, the signal multiplied by i, is fed through the OFFT and the real part is measured. The first cycle measures the real part of the result and the second measures the complex part.

Having both the real and imaginary part of the FFT at the output allows the OFFT to be reused in multiple cycles to generate the 2D FFT required for convolutional neural networks. Additionally, the 1D Cooley-Tukey algorithm can be divided over multiple cycles due to its recursive nature. This allows the size of the OFFT to be scaled appropriately for the application, with a trade between the number of ADCs and DACs and their operating speed. That is, a large number of ADCs and DACs can be replaced by a smaller number operating at a higher speed.

The choice of allocating delay presents another design decision. The architecture can be serial, with a single sample modulator (Fig. 4(a)) and optical delay, or parallel, with N sample modulators (Fig. 4(b)). While the serial case consumes less power, with only one DAC, it also has a lower convolution rate since only one convolution result can be produced within the period of the longest delay path.

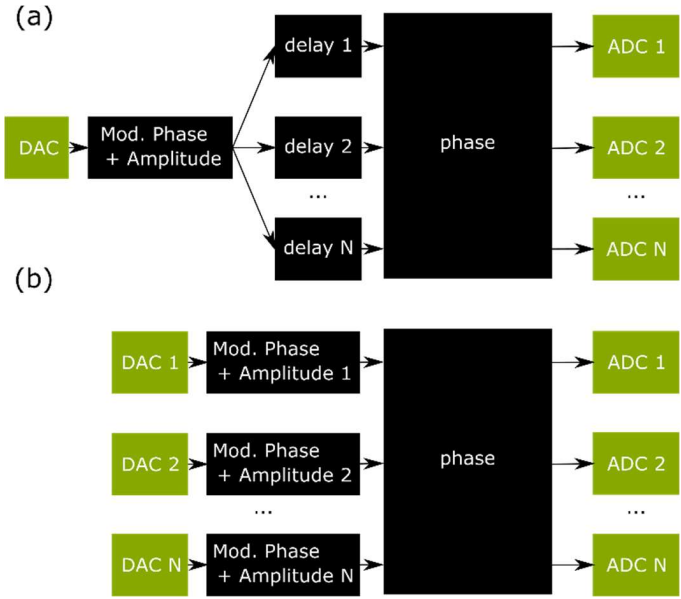

Fig. 4. The optical convolution can be run in serial with a single DAC and optical delay in the photonic network, (a) or with N parallel DACs and no additional delay in the photonic network (b).

## V. POWER CONSUMPTION

It is apparent that due to the passive nature of the OFFT network, the primary power consumer in the small N OFFT is found in the conversion between the digital and analog domains. If the OFFT were directly connected to an analog fully connected neural network or some other analog processor, the only power consumed by the OFFT would come from optical propagation losses, phase compensation, and the coherent optical source. However, todays dominant computer architecture is digital, and to be practical, the OFFT implementation must interface to the digital domain. The power consumption analysis becomes an analysis of digital and analog conversions.

Estimating the power budget for the optical FFT alone, we here consider two cases; (i) taking the state-of-the-art foundry processes of Silicon photonics [7], and (ii) emerging nanophotonics solutions on the other [8], we can bound the power budget for the optical portion of the on-chip FFT as follows. The only active portion of the FFT is the sampling of the output of the FFT transfer function performed by electro-optic modulators (EOM). Their output is then converted by a photodetector (and a possible TIA) and ADCs back into an electronic signal. Hence the only active power consumed dissipates at the sampling EOMs, the Rx leakage current (if present), and the laser source feeding the OFFT, which we consider off-chip due to temperature instabilities and subsequent thermo-optical impact on the phase sensitive OFFT.

In contrast, if we assume a nanophotonics solution for the on-chip EOM and Rx (case (ii)), but keeping the laser off chip as before, the photonic (excluding ADCs, DACs, and processor) power consumption for N = 64 OFFT is reduced from 146 W to 0.29 W, a reduction of 510 times. The main power reduction in nanophotonics is gained at the modulator which can required sub 1V power levels improved by enhanced light-matter-interactions [9] and unity-strong index changes shown by emerging EO active materials such as transparent conductive oxides (TCO) [10] and 2D dimensional materials such as Graphene's sensitive [11]. However, nanolasers or nano-LEDs could be considered on-chip in future improvements, which have higher conversion ratios compared to bulky sources enabled by their high spontaneous emission factors (i.e. high optical pump efficiencies and relative faster each of the lasing threshold [12]. In addition, plasmonic-silicon hybrid integration schemes have been predicted [13] to deliver higher performance for communication compared to either electronic, plasmonic, or silicon-photonic solutions alone, since the best of both worlds can be harnessed.

Modeling the power vs performance characteristics of the OFFT using the highest speed DAC and ADC found in literature today, and comparing against NVIDIA P100 GPU, shows up to four orders of magnitude better performance even with the high power consumed by the ADCs and DACs. The NIVIDIA P100 performs 1.6 TFLOPS during single precision 1024 length FFT [14]. Assuming the 1D FFT requires $2Nlog_2(N)$ multiplication operations and $3Nlog_2(N)$ addition operations there will be a total of $5Nlog_2(N)$ FLOPS per 1D FFT of length N. To generate a 2D FFT with an edge length of N from a 1D FFT there will be 2N 1D FFTs of length N for each 2D FFT. Each convolution requires one forward FFT, one $N^2$ multiplication, and one inverse FFT. Then the number of FLOPS to convolutions becomes $20N^2log_2(N) + N^2$. With the NVIDIA P100 this results in a convolution rate of 7 KHz with N = 1024 and 150 KHz with N = 256.

The highest speed DAC in literature operates at 100 GSa/s and consumes 2.5 W [15]. The highest speed ADC available today is the Fujitsu CHAIS 56 GSa/s and consumes 2 W per channel [16]. Assuming integrated germanium photodetectors [17] with a reverse bias of 8 V and 250 µW of optical power, the power consumption in each photodiode is approximated as 2.4 µW.

Using these assumptions, we modeled both serial and parallel implementations of the OFFT convolutional architecture and compared them to the NVIDIA P100 GPU (Fig. 5) using a Figure of Merit (FoM) of convolutions s$^{-1}$ W$^{-1}$ m$^{-2}$. The results of the analysis show that with a small convolution size the photonic approach is more efficient than the P100. However, the advantage diminishes as the convolution scales. This is due to the approximate scaling of the FoM with $1/(N^4 \log_2 N \, 10^{(N \log_2 N)/10})$ in the serial photonic case, $1/(N^4 \log_2 N)$, in the parallel photonic case, and $1/(N^2 \log_2 N)$ in the electronic case. The contribution to the serial photonic FFT scaling are from $1/(N \log_2 N)$ in area, $1/(10^{(N \log_2 N)/10})$ in optical power, $1/N$ in ADC power, $1/N^2$ in samples to area. Shortly beyond N = 10$^2$ the power efficiency of the P100 overtakes the serial photonic approach and near N = 10$^4$ the P100 passes the parallel OFFT architecture.

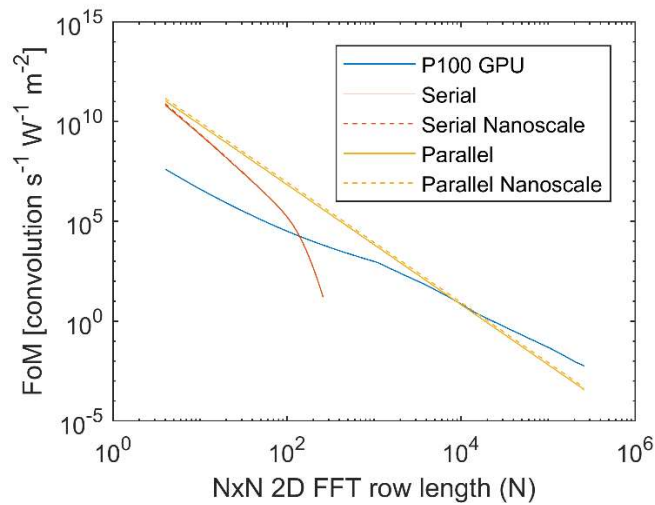

Fig. 5. A model of the figure of merit, Convolutions s$^{-1}$ W$^{-1}$m$^{-2}$, vs N shows the OFFT architecture outperforming the NVIDIA P100 in for N < 10$^2$ in a serial configuration and less than N < 10$^4$ in parallel configurations. The model assumes $20N^2 log_2(N) + N^2$ FLOPS per convolution, ADCs with 56 GSa/s @ 2 W, and DACs with a max sample rate of 100 GA/s @ 2.5 W. Photonics performance assumes IMEC Silicon Photonics ePIXfab optical performance spiral with a base optical loss of 0.686 dB, optical modulators with optical loss of 3.49 dB, 2x2 couplers with an optical loss of 0.991 dB, optical splitters with 0.5 dB insertion loss, and optical grating couplers with 4 dB loss. NVIDIA P100 GPU FLOPS performance taken from NVIDIA datasheet for single precision floating point.

## VI. CONCLUSION

The wave nature of coherent light allows for efficient arithmetic operations around the phase-amplitude plane. These operations can be harnessed to perform computations, including the FFT. The greatest power consumer in this type of computer is not in the arithmetic processing, which is almost negligible, but rather in the conversion from digital to analog, from electronic to photonic, from photonic to electronic, and from analog back to digital. Each of these conversions presents a significant hurdle to the power consumption of the architecture. As Moore's law reaches an end and future computer architectures continue to demand higher efficiencies, this mechanism for low-power computing may become an important supplement to digital computing.

ACKNOWLEDGEMENTS

V.S. is supported by Air Force Office of Scientific Research-Young Investigator Program under grant FA9550-14-1-0215, and under grant FA9550-15-1-0447.